\documentclass[12pt]{article}
\usepackage{graphicx} % Required for inserting images
\usepackage{amsfonts,amssymb,amsmath}
\usepackage{longtable,tabularx}
\usepackage[letterpaper,top=2.92cm,bottom=3.15cm,left=1.9cm,right=1.9cm,
marginparwidth=1.75cm,headheight=0.96cm,footskip=1.5cm]{geometry}
\usepackage{hyperref}
\usepackage{xcolor}
\usepackage[titletoc,title]{appendix}
\usepackage{subcaption}
\usepackage{mathtools}
\usepackage{authblk}

\usepackage[
backend=biber,
style=ieee,
]{biblatex}
\title{A non-history dependent temporal superposition algorithm for the finite line source solution}
 
\author[1]{Alberto Lazzarotto}
\author[1]{Marc Basquens}
\author[2]{Massimo Cimmino}
\affil[1]{Department of Energy Technology, Division of Applied Thermodynamics and Refrigeration, KTH Royal Institute of Technology, Stockholm, Sweden}
\affil[2]{Department of Mechanical Engineering, Polytechnique Montréal, Montréal, Canada}

\date{}

\begin{document}

\maketitle

\begin{abstract}
 Simulations of the operation of fields of borehole heat exchangers involve a wide spectrum of time scales, and hourly simulations for decades are required for the evaluation of the heat transfer in the subsurface due to these systems. Most current models rely on time and space superposition of fundamental analytical solutions of the heat equation to build the solution for complex borehole fields configurations and loading conditions. These procedures are robust and accurate but do not have favorable scaling properties, and the problems can become quickly computationally intractable as the size increases. In this context, acceleration algorithms for temporal superposition are key to overcome this limitation. 
   This paper presents developments on the so-called “non-history dependent" acceleration scheme and its application to the point and line source solutions, which are commonly used as building blocks in borehole field simulations. 

   The results obtained show promising properties as the computational complexity of the proposed algorithm is linear in the number of time steps, and near double precision accuracy can be achieved by refining the discretizations used to compute the integrals arising from the scheme.    
\end{abstract}

\section{Introduction}

Detailed thermal modeling of borehole fields is essential for the accurate design and operation of geothermal systems, ensuring that the expected performance is maintained throughout the lifespan of a shallow geothermal installation. Current modeling techniques primarily use “response factors", the method of images, and superposition of the effects as foundational tools to build solutions for the heat transfer problem in the subsurface \cite{eskilson1987, carslaw1959conduction, CIMMINO2014641}. Response factors represent how ground temperature changes over time due to a step heat source in an infinite volume. The method of images pairs a set of heat sources with a corresponding set of fictitious heat sources called “images" mutually symmetric with respect to a given plane, and enables representing heat transfer in a semi-infinite solid with boundary conditions that are relevant for geothermal applications. Finally, superposition allows to account for variations in heat distribution over time and space. A naïve implementation of time superposition results in a complexity of $\mathcal{O}(N_t^2)$, where $N_t$ is the number of uniformly spaced time steps in the simulations. This complexity is not suitable for conducting 20-year hourly simulations, which are typically required for designing borehole systems.

Several strategies have been proposed in the literature to reduce the computational complexity of the time superposition problem and accelerate simulations. Load aggregation methods \cite{Yavuzturk_Jeffrey, Bernier2004, ClaJav2012, MitSpi2019} that reduce the complexity by aggregating the effect of past loads in the time superposition evaluation were developed. Later, in \cite{MARCOTTE2008651}, the use of the Fast Fourier Transform (FFT) algorithm was proposed to compute the discrete convolution. Finally, the so-called “non-history dependent” method was ideated in \cite{lamarche2007}. This method is the least studied of the three techniques, but has theoretically good properties in terms of complexity and is a marching scheme, suitable for integration in borehole network simulation \cite{LAZZAROTTO2014265, LAMARCHE2017466} and building simulations. In this article, we focus on this method.

The original “non-history dependent” method studied the particular case of the infinite cylindrical source solution evaluated at the borehole radius and reduced the computational complexity of the time superposition process by leveraging its structure \cite{lamarche2007}. An extension of this scheme for generic responses by means of the Inverse Laplace Transform was put forward by Lamarche in \cite{lamarche2009}. However, in that work, the focus was put on numerical inversion of the Laplace Transform, which is in general  an ill-conditioned problem and, as such, can be unstable and dependent on the tuning of the numerical algorithm parameters.
In the present paper we explore the “non-history dependent” scheme using the analytical Inverse Laplace transform. 
Although this is hard to do in general, it can be done in the particular case of the point source solution \cite{ours}.
While the point source solution cannot be applied directly for geothermal calculations, it is the fundamental building block to build other relevant solutions for heat transfer in solids. In particular, this allows us to extend the “non-history dependent” method to responses to a finite line source which is the critical solution for practical applications in borehole field simulations \cite{CIMMINO2014641, LAMARCHE2007188, claesson2011analytical}. 

The paper is structured as follows. In Section \ref{section: methodology}, we review the original “non-history dependent” scheme in \cite{lamarche2007} and the generalization explained in \cite{lamarche2009}. Then, we apply it to derive a “non-history dependent” approach for the temperature field generated by a point source, and use the result to compute the temperature generated by a finite line source.
In Section \ref{section: integration}, we develop a numerical scheme to compute the integrals that appear in the expressions found in Section \ref{section: methodology}.
In Section \ref{section: numerical_analysis}, we analyze the precision of the algorithm and study its computational complexity.

\section{Methodology}
\label{section: methodology}
Lamarche and Beauchamp \cite{lamarche2007} derived a scheme that enables to evaluate the response to an infinite cylindrical source with time varying intensity, and advance a step forward in time using only information regarding the current time step. We provide a short outline of that method here.

The quantity of interest is the temperature at time $t$ caused by a heat extraction load $q^\prime(t)$ during the time interval $[0, t]$.
According to Duhamel's theorem, this is given by the convolution
\begin{align*}
    T(t) - T_0 = \frac{1}{k_g} \int_0^t  \, q^\prime(\tau) \frac{d{h}}{dt}(t-\tau) \ \mathrm{d}\tau \ ,
\end{align*}
where $k_g$ is the thermal conductivity, $q^\prime(t)$ is the load at time $t$, and $\frac{d{h}}{dt}$ the impulse response of the particular system on which the load is applied. It should be remarked that the units of $q^\prime(t)$ depend on what the units of $\frac{d h}{dt}$ are. In the paper we will deal with linear and point sources, hence $q^\prime(t)$ will be expressed in W/m and W, respectively. Note that $T$, $q^\prime$ and $\frac{d{h}}{dt}$ depend on spatial coordinates, however for the sake of notational simplicity we first address the case of a generic $\frac{d{h}}{dt}$ dropping the dependence on space, and then recover it when addressing specific impulse response functions, e.g. point source to point target. 
%Note that we omitted the spatial dependence from the previous expression for notational convenience.
Following \cite{lamarche2007}, we will use the non-dimensional time $\tilde{t} = \frac{\alpha}{r_b^2} t$ and the non-dimensional distance $\tilde{r} = \frac{r}{r_b}$ to rewrite this expression. Here, $r_b$ is the borehole radius and $\alpha$ the thermal diffusivity. We reparametrize $T$ with $t(\tilde{t})$ and change variable $\tilde{\tau} = \frac{\alpha}{r_b^2} \tau$ to obtain the expression

\begin{align*}
    T(\tilde{t}) - T_0 = \frac{1}{k_g} \frac{r_b^2}{\alpha} \int_0^{\tilde{t}} \, q^\prime \left(\tau(\tilde{\tau}) \right) \frac{d{h}}{dt} \left( t(\tilde{t})-\tau(\tilde{\tau}) \right) \ \mathrm{d}\tilde{\tau} \ ,
\end{align*}

where now $t$ and $\tau$ are to be understood as functions of $\tilde{t}$ and $\tilde{\tau}$, respectively. It is also useful to define $\frac{d{\tilde{h}}}{d\tilde{t}}(\tilde{t})$ as the reparametrized impulse response and $\tilde{q}^\prime(\tilde{t})$ as the reparametrized load with $t(\tilde{t})$,  so we can write
\begin{align}
    \label{eq: temperature_convolution}
    T(\tilde{t}) - T_0 = \frac{1}{k_g} \frac{r_b^2}{\alpha} \int_0^{\tilde{t}} \, \tilde{q}^\prime(\tilde{\tau}) \frac{d{\tilde{h}}}{d\tilde{t}}(\tilde{t}-\tilde{\tau}) \ \mathrm{d}\tilde{\tau} \ .
\end{align}

The key insight of the scheme was to realize that the impulse response of the infinite cylindrical source $\frac{d\tilde{h}_\text{ICS}}{d\tilde{t}}$ \cite{lamarche2009}, due to its particular form, could be expressed in the form
\begin{align}
\label{eq: impulse_exponential_form}
    \frac{d\tilde{h}_\text{ICS}}{d\tilde{t}}(\tilde{t}) = \int_0^\infty e^{-\zeta^2 \tilde{t}} v(\zeta) \ \mathrm{d}\zeta \ ,
\end{align}
for a certain function $v(\zeta)$. 

Substituting Equation \eqref{eq: impulse_exponential_form} into Equation \eqref{eq: temperature_convolution} and changing the order of integration between the variables $\tilde{\tau}$ and $\zeta$ allow to rewrite the temperature in Equation \eqref{eq: temperature_convolution} at a given time as
\begin{align}
\label{eq: temperature_from_F}
     T(\tilde{t}) - T_0 = \frac{1}{k_g} \frac{r_b^2}{\alpha} \int_0^\infty F \left(\zeta, \tilde{t} \right) \ \mathrm{d}\zeta \ ,
\end{align}
where
\begin{align}
\label{eq: F_definition}
    F(\zeta, \tilde{t}) = \int_0^{\tilde{t}} \, e^{-\zeta^2 (\tilde{t}-\tilde{\tau})} \tilde{q}^\prime(\tilde{\tau}) v(\zeta) \ \mathrm{d} \tilde{\tau}  \ .
\end{align}
Note that $F(\zeta, 0) = 0$.
The crucial property of $F$ is the following:
\begin{align}
\begin{split}    
\label{eq: F_recursive_general}
    F(\zeta, \tilde{t}+\Delta\tilde{t}) &= \int_0^{\tilde{t}} \, e^{-\zeta^2 (\tilde{t}+\Delta\tilde{t}-\tilde{\tau})} \tilde{q}^\prime(\tilde{\tau}) v(\zeta) \ \mathrm{d} \tilde{\tau} 
    \\
    &\hspace{10pt} +\int_{\tilde{t}}^{\tilde{t}+\Delta\tilde{t}} \, e^{-\zeta^2 (\tilde{t}+\Delta\tilde{t}-\tilde{\tau})} \tilde{q}^\prime(\tilde{\tau}) v(\zeta) \ \mathrm{d} \tilde{\tau} \\ 
    &=  e^{-\zeta^2 \Delta\tilde{t}} F(\zeta, \tilde{t}) + \int_{\tilde{t}}^{\tilde{t}+\Delta\tilde{t}} \, e^{-\zeta^2 (\tilde{t}+\Delta\tilde{t}-\tilde{\tau})} \tilde{q}^\prime(\tilde{\tau}) v(\zeta) \ \mathrm{d} \tilde{\tau} \ ,
\end{split}
\end{align}
namely, $F$ at time $\tilde{t}+\Delta\tilde{t}$ depends on itself at a previous time $\tilde{t}$ and the load applied during the time interval $[\tilde{t},\tilde{t}+\Delta\tilde{t}]$ in between the two times.

Equation \eqref{eq: F_recursive_general} becomes specially simple if we assume that the load $\tilde{q}^\prime$ is a piece-wise constant function %discretized in time (i.e., a sum of step functions in $\tilde{t}$) 
since the remaining integral can be evaluated. This yields a simple recursive formula to update $F$ to the next time step:
\begin{align}
\label{eq: F_recursive}
    F(\zeta, \tilde{t}+\Delta \tilde{t}) = e^{-\zeta^2 \Delta \tilde{t}} F(\zeta, \tilde{t}) + \tilde{q}^\prime(\tilde{t}) \left( 1-e^{-\zeta^2 \Delta \tilde{t}} \right) \frac{v(\zeta)}{\zeta^2} \ ,
\end{align}
which only depends on $F$ at the previous time step and the last load applied to the system. The temperature is then obtained by integrating $F$ as in Equation \eqref{eq: temperature_from_F}.

Using the fact that at the initial time $F$ vanishes, Equation \eqref{eq: F_recursive} can be further manipulated to highlight that $F(\zeta, \tilde{t})$ is the product of two terms: a time dependent term $\Hat{F}(\zeta, \tilde{t})$ which accounts for the load history and satisfies the recurrence relation
\begin{align}
    \label{eq: F_recursive_variable_separation_1}
    \Hat{F}(\zeta, \tilde{t}+\Delta \tilde{t}) =  e^{-\zeta^2 \Delta \tilde{t}} \Hat{F}(\zeta,\tilde{t}) + \tilde{q}^\prime(\tilde{t}) \left(1-e^{-\zeta^2 \Delta \tilde{t}} \right) \ ,
\end{align}
and a configuration term, which accounts for the geometry of the heat source and temperature evaluation region, given by
\begin{align*}
    \frac{v(\zeta)}{\zeta^2} \ .
\end{align*}

Then, one can write the original $F$ as
\begin{align}
\label{eq: F_recursive_variable_separation_2}    
    F(\zeta, \tilde{t}+\Delta \tilde{t})   
    % =  \Bar{F}(\zeta,\tilde{t} +\Delta \tilde{t}) \,  \frac{v(\zeta)}{\zeta^2}   
    = \Hat{F}(\zeta,\tilde{t} +\Delta \tilde{t}) \,  \frac{v(\zeta)}{\zeta^2} \ .
\end{align} 

Although this method relies on the very particular form of Equation \eqref{eq: impulse_exponential_form} of the impulse response of the infinite cylindrical source, Lamarche points out in \cite{lamarche2009} that generic impulse responses can be brought into this form, making them eligible for the method. The idea is as follows.
We would like to rewrite any impulse response $\frac{d\tilde{h}}{d\tilde{t}}$ as
\begin{align}
\label{eq: impulse_exponential_form_general}
    \frac{d\tilde{h}}{d\tilde{t}}(\tilde{t}) = \int_0^\infty e^{-\zeta^2 \tilde{t}} v(\zeta) \ \mathrm{d}\zeta \ ,
\end{align}
for an unknown function $v$. Note that by performing the change of variable $p=\zeta^2$, we obtain the expression
\begin{align*}
     \frac{d\tilde{h}}{d\tilde{t}}(\tilde{t}) = \int_0^\infty e^{-p \tilde{t}} \frac{v(\sqrt{p})}{2 \sqrt{p}} \ \mathrm{d}p \ = \mathcal{L}\left\{ \frac{v(\sqrt{p})}{2 \sqrt{p}} \right\}(\tilde{t}) \ ,
\end{align*}
which is the Laplace transform of the function $\frac{v(\sqrt{p})}{2 \sqrt{p}}$. Then, one can find the required $v$ by means of the inverse Laplace transform: 
\begin{align}
\label{eq: laplace_link}
    v(\zeta) = 2 \zeta \mathcal{L}^{-1} \left\{ \frac{d\tilde{h}}{d\tilde{t}}(\tilde{t})\right\} (\zeta^2)\ .
\end{align}

\begin{figure*}[!htb]
    \centering
     \includegraphics{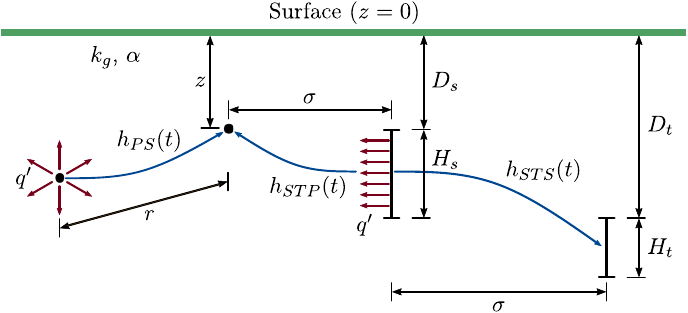}
    \caption{Geometry of the three considered heat source solutions.}
    \label{fig: source-geometries}
\end{figure*}

Although this is a powerful method, in practice, it might be difficult to numerically compute the expression in Equation \eqref{eq: laplace_link} for a particular system of interest. 
The function $v$ may be a highly oscillatory function, which can make the problem numerically untreatable.
%This might be true even numerically, as the inverse Laplace transform may behave unstably. 
This is, unfortunately, the case of the finite line source, most relevant for practical purposes. However, a closed form solution for the inverse Laplace transform of the point source impulse response exists and, as we discuss in Section \ref{section: segment_to_point}, it is possible to use it to indirectly compute the inverse Laplace transform of the finite line source, enabling us to use the iterative method. Figure \ref{fig: source-geometries} shows the geometry of the problem for the three considered heat source solutions: the point source solution ($h_{PS}$), the finite line segment to point solution ($h_{STP}$), and the finite line segment to segment solution ($h_{STS}$).

\subsection{Point source solution}
\label{section: point_source}

We start by considering the fundamental solution, which, even if it has limited application on its own, is the building block that allows to build solutions for a large number of system configurations relevant in several applications.

The impulse response of the point source solution in a homogeneous and isotropic medium is
\begin{align}
\label{eq: point_source_expression}
    \frac{dh_{PS}}{d t}(t) \vcentcolon =&  \frac{\alpha}{(4 \pi \alpha)^{3/2}}\,\frac{1}{t^{\;3/2}} e^{ -\frac{r^2}{4 \alpha t}  } \\
    \nonumber
    &=  \, \frac{\alpha}{r_b^3} \, \frac{1}{(4 \pi )^{3/2}}\,\frac{1}{\tilde{t}^{\;3/2}} e^{ -\frac{\tilde{r}^2}{4 \tilde{t}}  }  = \vcentcolon  \frac{d\tilde{h}_{PS}}{d \tilde{t}}(\tilde{t}) \ ,
\end{align}
where $\alpha$ is the thermal diffusivity and $k_g$ is the thermal conductivity of the medium.
As discussed above, we write it in terms of the non-dimensional time and distance.

We first need to compute the function $v$ by using Equation \eqref{eq: laplace_link}.
The relevant inverse Laplace transform in this case is \cite{oberhettinger2012tables}
\begin{align}\label{eq: inverse_laplace_expression}
    \mathcal{L}^{-1} \left\{s^{-3/2} e^{-a^2/s} \right\} (p) =
    \frac{1}{\sqrt{\pi} a} \, \sin(2 a p^{1/2}) \ .
\end{align}

Then, using Equations \ref{eq: laplace_link}, \ref{eq: point_source_expression}, and  \ref{eq: inverse_laplace_expression} one finds that the function $v_{PS}(\zeta, \tilde{r})$ for the point source is given by
% \begin{align*}
%     \frac{d\tilde{h}}{d \tilde{t}}(\tilde{t}) = \frac{\alpha}{k_g} \, \frac{1}{2 \pi^2 r_b^3 \tilde{r}} \,\int_0^\infty e^{-\zeta^2 \tilde{t}} \zeta \, \sin(\tilde{r}\zeta) \ \mathrm{d}\zeta \ .
% \end{align*}
\begin{align} 
\label{eq: u_point_source}
    \frac{v_{PS}(\zeta, \tilde{r})}{\zeta^2} = \frac{\alpha}{2 \pi^2 r_b^3 } \, \frac{\sin(\tilde{r}\zeta)}{\tilde{r} \zeta} \ .
\end{align}

Note that the dependency on $\tilde{r}$ was reintroduced to emphasize the fact that, for the point source, the configuration function $v_{PS}(\zeta, \tilde{r})$ depends on the distance between the source and evaluation points.
Given $v_{PS}(\zeta, \tilde{r})$, the temperature $T(\tilde{t})$ can be computed using Equations \ref{eq: temperature_from_F}, \ref{eq: F_recursive_variable_separation_1}, \ref{eq: F_recursive_variable_separation_2}.

% Then, the function $F$ is determined by the recursive relation \eqref{eq: F_recursive}:
% \begin{align}
% \label{eq: F_point_source}
%     F(\zeta, \tilde{t}+\Delta \tilde{t}) = e^{-\zeta^2 \Delta \tilde{t}} F(\zeta, \tilde{t}) + \frac{\alpha}{k_g} \, \frac{1}{2 \pi^2 r_b^3 \tilde{r}} \tilde{q}^\prime(\tilde{t}) \left(1 - e^{-\zeta^2 \Delta \tilde{t}} \right) \frac{\sin{\tilde{r} \zeta}}{\zeta} \ ,
% \end{align}
% and $F(z, 0) = 0$.
% Since the initial value of $F$ vanishes, we can define a new function $\tilde{F}$ by $F(\zeta, \tilde{t}) = \frac{1}{2 \pi^2 k_g r_b} \tilde{F}(\zeta, \tilde{t})$ and alternatively write an updating rule for $\tilde{F}$:
% \begin{align}
% \label{eq: F_point_source_final}
%     \tilde{F}(\zeta, \tilde{t}+\Delta \tilde{t}) = e^{-\zeta^2 \Delta \tilde{t}} \tilde{F}(\zeta, \tilde{t}) + \tilde{q}^\prime(\tilde{t}) \left(1 - e^{-\zeta^2 \Delta \tilde{t}} \right) \frac{\sin{\tilde{r} \zeta}}{\tilde{r}\zeta} \, ,
% \end{align}
% with $ \tilde{F}(\zeta, 0) = 0$. Finally, the temperature at each time step is given by
% \begin{align}
% \label{eq: temperature_from_F_final}
%     T(\tilde{t}) - T_0 = \frac{1}{k_g^2} \, \frac{1}{2 \pi^2 r_b} \int_0^\infty \tilde{F}(\zeta, \tilde{t}) \ \mathrm{d}\zeta \ .
% \end{align}

\subsection{Response at a point due to a uniform load on a finite line}
\label{section: segment_to_point}

The impulse response of a segment source with an extreme located at $(x^\prime,y^\prime,z^\prime = D)$ of length $H$ evaluated at a point $(x, y, z)$ can be readily obtained by integrating the fundamental solution along the interval $z \in [D, D+H]$
\begin{align*}
     \frac{d\tilde{h}_{STP}}{d \tilde{t}}(\tilde{t}) &= \int_D^{D+H} \frac{\alpha}{r_b^3} \, \frac{1}{(4 \pi )^{3/2}}\,\frac{1}{\tilde{t}^{\;3/2}} e^{ -\frac{\sigma^2 + (z-z^\prime)^2}{4 r_b^2 \tilde{t}} } \ \mathrm{d}z^\prime \\
     &=  \frac{\alpha}{8\pi r_b^2 \tilde{t}} \, e^{-\frac{\tilde{\sigma}^2}{4 \tilde{t}}} \left( \text{erf}\left( \frac{z-D}{2 r_b \sqrt{\tilde{t}}} \right) - \text{erf}\left( \frac{z-D-H}{2r_b \sqrt{\tilde{t}}} \right)\right) \ ,
\end{align*}
where $\sigma = \sqrt{(x-x^\prime)^2+(y-y^\prime)^2}$, and $\tilde{\sigma} = \sigma / r_b$.
However, it is not easy to compute the inverse Laplace transform of this expression, and hence, to find a suitable expression for the iterative scheme.

An alternative solution is to leverage that we know the function $v$ for the fundamental solution to build $v$ for the finite line source to a point:
% We briefly restore here the dependency on the spatial variables for clarity.
\begin{align*}
    \frac{d\tilde{h}_{STP}}{d \tilde{t}}(\tilde{t}) &= \int_D^{D+H} \frac{d\tilde{h}_{PS}}{d \tilde{t}}(\tilde{t}, \tilde{r}(z, z^\prime))  \ \mathrm{d}z^\prime \\
    &=  \int_D^{D+H} \int_0^\infty e^{-\zeta^2 \tilde{t}} v_{PS}\left(\zeta, \tilde{r}(z, z^\prime)\right) \ \mathrm{d}\zeta \ \mathrm{d}z^\prime \\
    &= \int_0^\infty e^{-\zeta^2 \tilde{t}}  \int_D^{D+H} v_{PS}(\zeta, \tilde{r}(z, z^\prime)) \ \mathrm{d}z^\prime \ \mathrm{d}\zeta  \\
    &=\vcentcolon  \int_0^\infty e^{-\zeta^2 \tilde{t}}   v_{STP}(\zeta, z)  \ \mathrm{d}\zeta \ .
\end{align*}
% so it is enough to integrate $v$ in $z^\prime$ to find $v_{FLS}$. 
% \begin{align} 
% \label{eq: v_FLS}
%     v_{FLS}(\zeta) =  \frac{\alpha}{k_g} \, \frac{1}{2 \pi^2 r_b^3 } \, \int_D^{D+H} \, \zeta \, \frac{\sin\left(\tilde{r}(z,z^\prime) \, \zeta\right)}{\tilde{r}(z,z^\prime)} \ \mathrm{d} z^\prime  ,
% \end{align}
Given this result, it is possible to get a closed expression for the finite line source geometry term $v_{STP}$ by simply integrating the function $v$ for a point source along the line of interest
\begin{align} 
\label{eq: u_FLS}
    \frac{v_{STP}(\zeta)}{\zeta^2} = \frac{\alpha}{2 \pi^2 r_b^3 } \, \int_D^{D+H} \, \frac{\sin\left(\tilde{r}(z,z^\prime) \, \zeta\right)}{\tilde{r}(z,z^\prime) \zeta} \ \mathrm{d} z^\prime  \ .
\end{align}

% Similarly to the point source case it is possible to write an iterative scheme to update $\tilde{F}_{FLS}$
% % It is A good strategy for the iterative scheme is to update $\tilde{F}_{FLS}$ just as in the point source case
% \begin{align*}
%     \tilde{F}_{FLS}(\zeta, \tilde{t}+\Delta \tilde{t}) = e^{-\zeta^2 \Delta \tilde{t}} \tilde{F}_{FLS}(\zeta, \tilde{t}) + \tilde{q}^\prime(\tilde{t}) \left(1 - e^{-\zeta^2 \Delta \tilde{t}} \right) \int_D^{D+H} \frac{\sin{\tilde{r} \zeta}}{\tilde{r}\zeta} \, \mathrm{d}z^\prime \, ,
% \end{align*}

% with $ \tilde{F}_{FLS}(\zeta, 0) = 0$ and then perform the extra integral in $z^\prime$ when evaluating the temperature at each time step
% \begin{align}
% \label{eq: temperature_from_F_final_fls}
%     T(\tilde{t}) - T_0 = \frac{1}{k_g^2} \, \frac{1}{2 \pi^2 r_b} \int_0^\infty \tilde{F}_{FLS}
%     %\int_D^{D+H} \tilde{F}_{FLS}
%     (\zeta, \tilde{t}) \ \mathrm{d}z^\prime \ \mathrm{d}\zeta \ .
% \end{align}

\subsection{Mean response at a finite line due to a uniform load on a finite line}
\label{section: segment_to_segment}

The same procedure as in section \ref{section: segment_to_point} can be applied to find the mean response on a target segment (starting at $z=D_t$ and of length $H_t$) due to the effect of the load on a source segment (starting at $z=D_s$ and of length $H_s$). 
In this case, the relation between impulse responses is
\begin{align*}
     \frac{d\tilde{h}_{STS}}{d \tilde{t}}(\tilde{t}) = \frac{1}{H_t} \int_{D_t}^{D_t+H_t} \int_{D_s}^{D_s+H_s} \frac{d\tilde{h}_{PS}}{d \tilde{t}}(\tilde{t}, \tilde{r}(z, z^\prime)) \ \mathrm{d}z^\prime \ \mathrm{d} z \ .
\end{align*}
Then, the geometry term $v_{STS}$ of the mean temperature along a segment due to a source placed on another segment is 
\begin{align} 
\label{eq: u_STS}
    \frac{v_{STS}(\zeta)}{\zeta^2} =  \frac{1}{H_t} \, \frac{\alpha}{2 \pi^2 r_b^3 } \, \int_{D_t}^{D_t+H_t} \int_{D_s}^{D_s+H_s}  \, \frac{\sin\left(\tilde{r}(z,z^\prime) \, \zeta\right)}{\tilde{r}(z,z^\prime) \zeta} \ \mathrm{d} z^\prime \, \mathrm{d} z  \ .
\end{align}

% Hence, the temperature at each time step can be computed by
% \begin{align}
% \label{eq: temperature_from_F_final_sts}
%     T(\tilde{t}) - T_0 = \frac{1}{k_g^2} \, \frac{1}{2 \pi^2 r_b} \int_0^\infty \int_{D_t}^{D_t+H_t} \int_{D_s}^{D_s+H_s} \tilde{F}_{STS}(\zeta, \tilde{t}) \ \mathrm{d}z^\prime \ \mathrm{d} z  \ \mathrm{d}\zeta \ ,
% \end{align}
% where $\tilde{F}_{STS}$ is updated as in the previous cases.

\section{Numerical Integration}
\label{section: integration}

Regardless of the scenario we are interested in, to be able to obtain the temperature with this method, one must integrate $F$ at each time step according to Equation \eqref{eq: temperature_from_F}. The integrand function $F$ is defined by Equation \eqref{eq: F_recursive_variable_separation_2} and is the product of a time dependent function $\Hat{F}$ and a configuration dependent function $v$. The latter one, for all three cases introduced in Section \ref{section: methodology}, is oscillating because of the term $\sin(\tilde{r} \zeta)$. 
% Regardless of the scenario we are interested into, in order to obtain the temperature with this method, we need to integrate $\tilde{F}$ at each time step according to the corresponding equation (\eqref{eq: temperature_from_F_final}, \eqref{eq: temperature_from_F_final_fls}, or \eqref{eq: temperature_from_F_final_sts}).
% The integrand in all three cases is defined by \eqref{eq: F_point_source_final}, and it is oscillating because of the term $\sin(\tilde{r} \zeta)$. 
The frequency of oscillation is the non-dimensional distance $\tilde{r}$ and, as a result, evaluations of the response far away from the source result in very fast oscillations of the integrand function.
Standard quadrature techniques such as Gauss integration may yield poor results in these cases and it is therefore necessary to use specialized methods for oscillating integrals.
In this work, we propose an adhoc scheme based on such techniques to perform the integrals.

We start by outlining a procedure for the point source in order to directly show the oscillatory term in the expression, but the strategy can be applied for all cases. By defining the constant
\begin{align*}
    C =  \frac{\alpha}{2 \pi^2 r_b^3 } \ ,
\end{align*}
the integrand function $F$ for the point source case can be written as 
\begin{align}
\label{eq: F_explicit_steps}
    F(\zeta, (n+1) \Delta \tilde{t}) =  C \, \left(1 - e^{-\zeta^2 \Delta \tilde{t}} \right) \frac{\sin{\tilde{r} \zeta}}{\tilde{r}\zeta} \sum_{i=0}^n  \left( \tilde{q}^\prime(i \Delta \tilde{t}) e^{-\zeta^2 (n-i) \Delta \tilde{t}} \right) \ .
\end{align}
% and focus on the integration of the function $\tilde{F}(\zeta, (n+1) \Delta \tilde{t}) = \frac{F(\zeta, (n+1) \Delta \tilde{t})}{C}$ ,

% By inspecting the function $\tilde{F}$ in equation \eqref{eq: F_point_source_final}, one can see that it can be written as
% \begin{align}
% \label{eq: F_explicit_steps}
%     \tilde{F}(\zeta, (n+1) \Delta \tilde{t}) = \left(1 - e^{-\zeta^2 \Delta \tilde{t}} \right) \frac{\sin{\tilde{r} \zeta}}{\tilde{r}\zeta} \sum_{i=0}^n  \left( \tilde{q}^\prime(i \Delta \tilde{t}) e^{-\zeta^2 (n-i) \Delta \tilde{t}} \right) \ .
% \end{align}

% It is very important to note that, although it is possible to analytically integrate expression \eqref{eq: F_explicit_steps} (at least in $\zeta$), we cannot do so in order to use the present method. This is because the mentioned expression contains a convolution (namely, between $q(\tilde{t})$ and $e^{-\zeta^2 \tilde{t}}$), which we are actively trying to avoid by devising an iterative algorithm. Hence, in order to leverage the computational advantages of the iterative method, we \textit{must} integrate $\tilde{F}$ numerically at each time step.
Note that, at any time step, all the terms are proportional to $e^{-\zeta^2 \Delta \tilde{t}}$ except for one, given by the last term of the sum.
Then, $F$ can be separated into two terms:
\begin{align}
\label{eq: F_r}
    F_\text{s}(\zeta, (n+1) \Delta \tilde{t}) &=  \tilde{q}^\prime (n \Delta \tilde{t}) \frac{v(\zeta)}{\zeta^2} = C \, \tilde{q}^\prime (n \Delta \tilde{t})  \frac{\sin{\tilde{r} \zeta}}{\tilde{r}\zeta} \ , \\
\label{eq: F_exp}
    F_\text{exp}(\zeta, (n+1) \Delta \tilde{t}) &= F(\zeta, (n+1) \Delta \tilde{t}) - F_\text{s}(\zeta, (n+1) \Delta \tilde{t}) \ ,
\end{align}
where $F_\text{exp}$ is proportional to $e^{-\zeta^2 \Delta \tilde{t}}$ and $F_\text{s}$ is not.
Note that both terms change at each time step, but while $F_\text{exp}$ grows in the number of summands, $F_\text{s}$ is always a single term. 
We will integrate each term separately and the desired result of the integral of Equation \eqref{eq: temperature_from_F} is obtained by adding their contributions.

As a remark, note that in general, $F_\text{s}$ can be interpreted as the term corresponding to the steady state of the unit response. To see this, it is enough to consider a unit constant load $q^\prime(\tilde{t}) = 1$ and take the limit $t^\prime \to \infty$ in Equation \eqref{eq: F_recursive}, which yields
\begin{align*}
    F(\zeta, \tilde{t} + \Delta\tilde{t}) = e^{-\zeta^2 \Delta\tilde{t}} F(\zeta, \tilde{t}) + v(\zeta) \frac{1 - e^{-\zeta^2 \Delta\tilde{t}}}{\zeta^2}  \xrightarrow[]{\Delta\tilde{t} \to \infty} \frac{v(\zeta)}{\zeta^2} \ , 
\end{align*}
which implies 
\begin{align}
\label{eq:limit_F_v}
    \lim_{\tilde{t} \rightarrow \infty} F(\zeta, \tilde{t}) = \frac{v(\zeta)}{\zeta^2} \ .
\end{align}
On the other hand, by Equations \eqref{eq: temperature_convolution} and \eqref{eq: temperature_from_F}, we have
\begin{align}
\label{eq:step_response_F}
    \int_0^\infty F(\zeta, \tilde{t}) \ \mathrm{d}\zeta = \int_0^{\tilde{t}} q^\prime(\tilde{\tau}) \frac{d{\tilde{h}}}{d\tilde{t}}(\tilde{t}-\tilde{\tau})  \ \mathrm{d}\tilde{\tau} = \int_0^{\tilde{t}} \frac{d{\tilde{h}}}{d\tilde{t}}(\tilde{t}-\tilde{\tau}) \ \mathrm{d}\tilde{\tau} = \tilde{h}(\tilde{t})  \ ,
\end{align}
since $\tilde{h}(0) = 0$.
By taking the limit $\tilde{t} \rightarrow \infty$ in Equation \eqref{eq:step_response_F}, and using Equation \eqref{eq:limit_F_v}, 
we find
\begin{align*}
    \int_0^\infty \frac{v(\zeta)}{\zeta^2} \ \mathrm{d}\zeta = \tilde{h}^\star \ , 
\end{align*}
where $\tilde{h}^\star$ denotes the steady state unit response. Therefore, by Equation \eqref{eq: F_r}, we conclude that
\begin{align*}
    \int_0^\infty  F_\text{s}(\zeta, (n+1) \Delta \tilde{t}) \ \mathrm{d}\zeta = \tilde{q}^\prime (n \Delta \tilde{t})  \tilde{h}^\star \ .
\end{align*}

% \begin{align}
% \label{eq: F_r}
%     \tilde{F}_\text{r}(\zeta, (n+1) \Delta \tilde{t}) &= \tilde{q}^\prime (n \Delta \tilde{t}) \frac{\sin{\tilde{r} \zeta}}{\tilde{r}\zeta} \ , \\
% \nonumber
%     \tilde{F}_\text{exp}(\zeta, (n+1) \Delta \tilde{t}) &= \tilde{F}(\zeta, (n+1) \Delta \tilde{t}) - \tilde{F}_\text{r}(\zeta, (n+1) \Delta \tilde{t}) \ ,
% \end{align}
% where $\tilde{F}_\text{exp}$ is proportional to $e^{-\zeta^2 \Delta \tilde{t}}$ and $\tilde{F}_\text{r}$ is not. Note that both terms change at each time step, but while $\tilde{F}_\text{exp}$ grows in the number of terms, $\tilde{F}_\text{r}$ is always a single term. We will integrate each term separately and the desired result of the integral of \eqref{eq: F_point_source_final} is obtained by adding their contributions.

\subsection{Integration of \texorpdfstring{$F_\text{s}$}{Fs}}

The explicit expression of $F_\text{s}$ for any time step is Equation \eqref{eq: F_r}, and in our cases of interest, we will be able to perform the integral analytically.
Note that except for the load at the current time step, the result will be independent of time.

In the case of the point source (subsection \ref{section: point_source}), we only need to integrate in $\zeta$ to obtain
\begin{align*}
    \int_0^\infty F_\text{s}(\zeta, (n+1)\Delta \tilde{t}) \ \mathrm{d}\zeta = C \, \tilde{q}^\prime(n\Delta \tilde{t}) \frac{\pi}{2} \frac{1}{\tilde{r}} \ .
\end{align*}

In the case of the finite line source (subsection \ref{section: segment_to_point}), there is a second integral in the variable $z^\prime$, yielding
\begin{align*}
    \int_D^{D+H} & \int_0^\infty F_\text{s}(\zeta, (n+1)\Delta \tilde{t}) \ \mathrm{d}\zeta \ \mathrm{d}z^\prime \\
    &= C \, \tilde{q}^\prime(n\Delta \tilde{t}) \frac{\pi}{2} \log{\left(\frac{z-D + \sqrt{\sigma^2 + (z-D)^2}}{z-D-H + \sqrt{\sigma^2 + (z-D-H)^2}}\right)} \ .
\end{align*}

In the case of the segment to segment response (subsection \ref{section: segment_to_segment}), there is yet a third integral
\begin{align}
\label{eq: seg_to_seg_F_r_integrated}
     \frac{1}{H_t} \int_{D_t}^{D_t+H_t} \int_{D_s}^{D_s+H_s} \int_0^\infty F_\text{s}(\zeta, (n+1)\Delta \tilde{t}) \ \mathrm{d}\zeta \ \mathrm{d}z^\prime \ \mathrm{d}z \ .
\end{align}
In order to express its result, let us define a few constants
\begin{align*}
    d_1 &= D_s + H_s - D_t \ , \\
    d_2 &= D_s - H_t - D_t \ , \\
    d_3 &= D_s - D_t \ , \\
    d_4 &= D_s + H_s - D_t - H_t \ , \\
    \beta(d) &= \sqrt{\sigma^2 + d^2} + d \log{\left(\sqrt{\sigma^2 + d^2}- d \right)} \ .
\end{align*}
Then, the result of the integral in Equation \eqref{eq: seg_to_seg_F_r_integrated} can be expressed as
\begin{align*}
    C \, \tilde{q}^\prime(n\Delta \tilde{t})\frac{\pi}{2 H_t} \left( \beta(d_4) + \beta(d_3) -\beta(d_1)-\beta(d_2)\right) \ .
\end{align*}

% As expected, this is the steady state of the mean temperature over a segment generated by a source segment \cite{sts_steady_state}.
The result obtained is consistent with the expression for the steady state of the mean temperature over a segment generated by a source segment presented in \cite{sts_steady_state}.

% \subsection{Integration of $\tilde{F}_\text{exp}$}
\subsection{Integration of \texorpdfstring{$F_\text{exp}$}{Fexp}}

It will be convenient to introduce a new function $\Bar{F}$ defined by
% \begin{align*}
%     F_\text{exp}(\zeta, \tilde{t}) = \, \Bar{F}(\zeta, \tilde{t}) \sin{\tilde{r} \zeta}\ ,
% \end{align*}
\begin{align*}
    F_\text{exp}(\zeta, \tilde{t}) = C \, \Bar{F}(\zeta, \tilde{t}) \sin{\tilde{r} \zeta}\ ,
\end{align*}
which is non-oscillatory. The explicit expression of $\Bar{F}$ is 
\begin{align}
\label{eq: F_expanded}
    \Bar{F}(\zeta, \tilde{t}) =  \frac{1}{\tilde{r}\zeta}  \left(  \left(1 - e^{-\zeta^2 \Delta \tilde{t}} \right) \sum_{i=0}^{n-1}   \tilde{q}^\prime(i \Delta \tilde{t}) e^{-\zeta^2 (n-i) \Delta \tilde{t}} - e^{-\zeta^2 \Delta \tilde{t}} \tilde{q}^\prime(n \Delta \tilde{t}) \right) \ .
\end{align}
Note that it is enough to just update $\Bar{F}$ at each time step rather than $F$.

By construction $F_\text{exp}$ and $ \Bar{F}$ are proportional to $e^{- \zeta^2 \Delta \tilde{t}}$, it is clear that they decay quickly with the integration variable $\zeta$, and since it will be convenient to have an integral over a finite domain, we can make the approximation 
% \begin{align}
% \label{eq: truncated_integral_F_exp}
%     \int_0^\infty \tilde{F}_\text{exp}(\zeta, \tilde{t}) \ \mathrm{d}\zeta \approx \text{Im}\left( \int_0^b \Bar{F}(\zeta, \tilde{t}) e^{i \tilde{r} \zeta} \ \mathrm{d}\zeta  \right) \ ,
% \end{align}
\begin{align}
\label{eq: truncated_integral_F_exp}
    \int_0^\infty F_\text{exp}(\zeta, \tilde{t}) \ \mathrm{d}\zeta \approx C \, \text{Im}\left( \int_0^b \Bar{F}(\zeta, \tilde{t}) e^{i \tilde{r} \zeta} \ \mathrm{d}\zeta  \right) \ ,
\end{align}
for a choice of truncated interval $b$. In Appendix \ref{section: appendix_error}, we provide an estimate of the error committed in doing so as a function of $b$.

We now need to evaluate an oscillatory integral in a closed interval. There are many methods available for such a problem and we chose a particular one proposed by Bakhvalov and Vasil'eva \cite{bakhvalov}, which has features that in our opinion are very suitable for this application. 
The method computes integrals of the form
\begin{align*}
    I(g, \omega) = \int_{-1}^1 g(x) e^{i \omega x} \ \mathrm{d}x \ ,
\end{align*}
by approximating $g$ by a truncated series up to $n$-th order of Legendre polynomials $P_k$ as
\begin{align*}
    g(x) = \sum_{k=0}^n c_k P_k(x) \ ,
\end{align*}
where 
\begin{align}
\label{eq: ck}
    c_k = \frac{2k+1}{2} \int_{-1}^1 g(x) P_k(x) \ \mathrm{d}x \ .
\end{align}
Then, by using that 
\begin{align*}
    \int_{-1}^{1} P_k(x) e^{i\omega x} \ \mathrm{d} x =  i^k \sqrt{\frac{2\pi}{\omega}} J_{k+\frac{1}{2}}(\omega) \ ,
\end{align*}
where $J_{\alpha}(x)$ is the Bessel function of the first kind of order $\alpha$, one can approximate the original integral as
\begin{align}
\label{eq: integral_approx}
    I(g, \omega) \approx \sqrt{\frac{\pi}{2\omega}} \, \sum_{k=0}^n \sum_{s=0}^{n} i^k (2k + 1) \, J_{k+\frac{1}{2}}(\omega) \, w_s P_k(x_s) g(x_s) \ ,
\end{align}
where $x_s, w_s$ are the nodes and weights, respectively, of the $(n+1)-$point Gauss-Legendre quadrature rule.
Note that the integral in Equation \eqref{eq: ck} has been also approximated by the Gauss-Legendre quadrature rule.
The accuracy of Equation \eqref{eq: integral_approx} depends heavily on the accuracy of the approximation of $g(x)$ via Legendre polynomials -- which can also be thought of as the fineness of the discretization of the interval induced by the Gauss-Legendre nodes that we use. 
While the original formulation of Bakhalov and Vasil'eva \cite{bakhvalov} is for integration in the interval $[-1,1]$, it can be easily extended for an arbitrary finite interval $[a,b]$. 
This is given by the formula \cite{evans1999}
\begin{align}
\label{eq: bakhalov}
    \int_a^b g(x) e^{i \omega x} \ \mathrm{d}x = m  e^{i \omega c} I ( G, m \omega) \ ,
\end{align}
where $G(x) = g(m x + c)$, $m = \frac{b-a}{2}$, and $c = \frac{b+a}{2}$. 

Note that a feature of this method, as reflected by Equation \eqref{eq: integral_approx}, is that the integrand, the dependency on the frequency $\omega$, and the dependency on the discretization (namely, $x_s, w_s$) are not intertwined. We can use this observation to define the matrix $P$ and the vectors $\Omega, f$ as
\begin{align*}
    P_{ks} = \sqrt{ \frac{\pi}{2} m} \, (2k + 1) w_s P_k(x_s) \ , \\
    \Omega_{k}(\omega) = i^k  \frac{e^{i \omega c}}{\omega^{1/2}} J_{k+\frac{1}{2}}(m \omega) \ ,  \\
    f_s = G(x_s) \ ,
\end{align*}
which allow us to rewrite Equation \eqref{eq: bakhalov} in matricial form
\begin{align}
\label{eq: integral_matricial}
    \int_a^b g(x) e^{i \omega x} \ \mathrm{d}x = \Omega^T(\omega) \, P \, f \ .
\end{align}

Due to the exponential nature of $\Bar{F}$, its value will be much larger around $0$ than around the cutoff $b$, and this effect is only amplified as we iterate through time due to the $e^{-\zeta^2 \Delta \tilde{t}}$ factor that is introduced at each time step, according to Equation \eqref{eq: F_explicit_steps}. Consequently, approximating $\Bar{F}$ by a low order polynomial on the interval $[0, b]$ may introduce a large error in the computation of the integral. 
Choosing a higher order for the Legendre series approximation may not be an efficient solution, since its points are distributed all along the integration interval, while the biggest contributions to the integral happen in small subintervals.
A balanced way to improve precision without dramatically increasing the number of points is to divide the interval $[0, b]$ into a number $m$ of smaller subintervals, possibly with different numbers of points $n$, taking into account where the steepest changes in $\Bar{F}$ happen. Then, Equation \eqref{eq: integral_matricial} should be used in each subinterval separately and all the results added together to obtain the desired result. This has the effect of refining the discretization in the regions where it is more beneficial in a more efficient way. 
\\

We are now ready to compute the relevant integrals.
In the case of the point source, we can directly use Equation \eqref{eq: integral_matricial} to obtain the integral of $\tilde{F}_\text{exp}$ via the approximation in Equation \eqref{eq: truncated_integral_F_exp}. 
However, since we have $\omega = \tilde{r}$, and our function to integrate Equation \eqref{eq: F_expanded} is proportional to $\frac{1}{\tilde{r}}$, we can modify the vector $\Omega(\omega)$ to 
\begin{align*}
    R_{\textit{PS}_k}(\tilde{r}) = i^k  \frac{e^{i \tilde{r} c}}{\tilde{r}^{3/2}} J_{k+\frac{1}{2}}(m \tilde{r}) \ , 
\end{align*}
to directly account for this extra factor. Then, we have

\begin{align}
\label{eq: integral_exp_point}
    \int_0^\infty F_\text{exp}(\zeta, \tilde{t}) \ \mathrm{d}\zeta  \approx C \, \text{Im}\left( R_\textit{PS}^T(\tilde{r}) \, P \, f \right) \ .
\end{align}

However, for the remaining cases, we also need to perform integrals in $z^\prime$ and $z$. There are some observations we can make:
\begin{itemize}
    \item The function resulting of the integral in $\zeta$ is no longer oscillatory. We can easily see this by integrating analytically a typical term of $\Bar{F}$:
    \begin{align*}
        \int_0^\infty e^{-n \zeta^2 \Delta \tilde{t}} \, \frac{\sin  \tilde{r} \zeta }{\tilde{r} \zeta} \ \mathrm{d} \zeta = \frac{\pi}{2\tilde{r}} \, \textrm{erf}\left( \frac{\tilde{r}}{2 \sqrt{n  \Delta \tilde{t}}}\right) \ ,
    \end{align*}
    which is clearly non-oscillatory. Therefore, we can use standard integration techniques for subsequent integrals.
    \item Both $z, z^\prime$ appear in the integrand only through the distance $\tilde{r}$. This means that only some terms in Equation \eqref{eq: bakhalov} are relevant for the integration, the rest being constant.
\end{itemize} 
Using these observations, we can easily perform the remaining integrals by slightly modifying Equation \eqref{eq: integral_matricial}. We will be using the standard Gauss-Legendre quadrature rule, hence we will need to evaluate the integrand at several distances $\tilde{r}_i$, to later multiply by the corresponding weight.

Let us start with the case of the response at a point due to a finite line source. Choose the number of points $N^\prime$ and denote by $z^\prime_i$ the nodes and by $v^\prime_i$ the weights of the $(N^\prime+1)-$point Gauss-Legendre quadrature rule.
Define the vector $R_\textit{STP}(v^\prime)$ by
\begin{align}
\label{eq: z_integral_discrete}
    R_{\textit{STP}_k}(v^\prime) &= \frac{H}{2} \sum_{i=0}^{N^\prime} R_{\textit{PS}_k}\left( \tilde{r}_i \right) v^\prime_i \ , \\
    \tilde{r}_i &= \frac{1}{r_b} \sqrt{\sigma^2 + (z - z^\prime_i)^2} \nonumber \ ,
\end{align}
where $\sigma$ is the distance in the $x,y$-plane between the segment and the evaluation point. Note that, effectively, computing Equation \eqref{eq: z_integral_discrete} is performing the integral in $z^\prime$ from Equation \eqref{eq: u_FLS}.
Then, we have
\begin{align}
\label{eq: integral_exp_seg_to_point}
    \int_D^{D+H} \int_0^\infty F_\text{exp}(\zeta, \tilde{t}) \ \mathrm{d}\zeta \ \mathrm{d} z^\prime \approx C \, \text{Im}\left( R_\textit{STP}(v^\prime)^T \, P \, f \right) \ .
\end{align}

For the segment to segment case, another modification is needed. Since there are two integrals, we need two sets of Gauss-Legendre quadrature nodes $z_i, z^\prime_i$ and weights $v_i, v^\prime_i$, of $(N+1)$ and $(N^\prime+1)$ points, respectively. Then, we define the vector $R_\textit{STS}(v, v^\prime)$ as
\begin{align}
    \label{eq: z_double_integral_discrete}
    R_{\textit{STS}_k}(v, v^\prime) &= \frac{H_s}{2} \frac{H_t}{2}\sum_{i=0}^N \sum_{j=0}^{N^\prime} R_{\textit{PS}_k}\left( \tilde{r}_{ij} \right) v_i v^\prime_j \ , \\ 
    \tilde{r}_{ij} &= \frac{1}{r_b} \sqrt{\sigma^2 + (z_i - z^\prime_j)^2} \nonumber \ .
\end{align}
Again, effectively, computing Equation \eqref{eq: z_double_integral_discrete} is performing the two integrals in $z, z^\prime$ from Equation \eqref{eq: u_STS}.
Finally, we have
\begin{align}
\label{eq: integral_exp_seg_to_seg}
    \frac{1}{H_t} \int_{D_t}^{D_t+H_t} \int_{D_s}^{D_s+H_s} &\int_0^\infty F_\text{exp}(\zeta, \tilde{t}) \ \mathrm{d}\zeta \ \mathrm{d} z^\prime  \ \mathrm{d} z \\
    &\approx \frac{1}{H_t} \, C \, \text{Im}\left( R_\textit{STS}(v, v^\prime)^T \, P \, f \right) \ .
\end{align}

We would like to remark that it is also possible to further refine the discretizations of one or both lines to increase precision effectively, similarly to what has been discussed before Equation \eqref{eq: integral_exp_point} for the integral in $\zeta$.

\section{Numerical analysis}
\label{section: numerical_analysis}
The goal of this section is to analyze both the precision and the computational complexity of the algorithms presented in Section \ref{section: integration}.

The accuracy of the algorithms depends on the choices that we make on $b$ and the discretizations both in $\zeta$ and in the line sources. In Appendix \ref{section: appendix_error}, we discuss the error introduced by the cutoff $b$, which can be brought below double precision for the right choice of $b$ depending on the rest of the parameters. Therefore, it remains to analyze how the error behaves as a function of the discretization.  

In order to measure the error, we compare the result with the convolution of the load with the impulse response. In practice, we evaluate the equivalent formula
\begin{align}
\label{eq: convolution}
    T(t) - T_0 &= \sum_{i=0}^{N_t} q^\prime \big((N_t-i) \Delta t \big) \frac{d h}{d t}\left(i \Delta t\right) \\
    \nonumber
    &= \sum_{i=0}^{N_t} \Big( q^\prime \big((N_t - i - 1) \Delta t \big) - q^\prime \big((N_t - i ) \Delta t\big) \Big) h\left(i \Delta t\right) \ ,
\end{align} 
using the step response.
The step response of the point source is given by
\begin{align*}
    h_{PS}(t, r) =\frac{\text{erfc}\left(\frac{r}{\sqrt{4 t \alpha}}\right)}{4\pi r k_g} \ .
\end{align*}
Although we do not have an analytical expression for the line to point $h_{STP}$ and line to line $h_{STS}$ step responses, we can obtain their numerical value by integrating $h_{PS}$ over the corresponding lines, analogously as discussed in Sections \ref{section: segment_to_point}, \ref{section: segment_to_segment}.
For the purpose of benchmarking, we compute Equation \eqref{eq: convolution} with a double accuracy precision $\varepsilon \simeq 2 \cdot 10^{-16}$.

\begin{figure*}[!htb]
    \centering
    \includegraphics{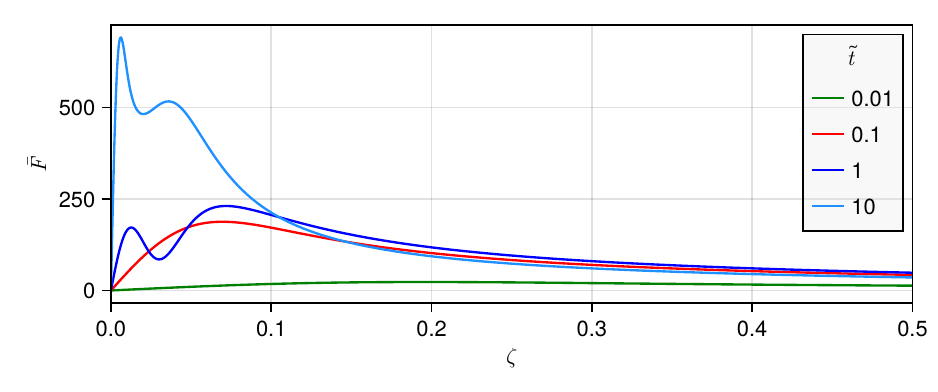}
    \caption{Function $\Bar{F}$ at different times for a synthetic hourly load with yearly and daily periodic oscillations.}
    \label{fig: F_interpolant}
\end{figure*}

As a representative example, we used the synthetic load
\begin{align}
\label{eq: synthetic_load}
    q^\prime(t) = 20 \sin{\left(\frac{2 \pi t}{8760}\right)} + 5\sin{\left(\frac{2 \pi t}{24}\right)} + 5 \ ,
\end{align}
during a period of $20$ years with hourly time steps to perform the simulations. This particular function aims to replicate the seasonal and daily variations in the load. Figure \ref{fig: F_interpolant} shows the $\Bar{F}$ function associated with this load at several points in time.

\begin{figure*}[!htb]
    \centering
    \includegraphics{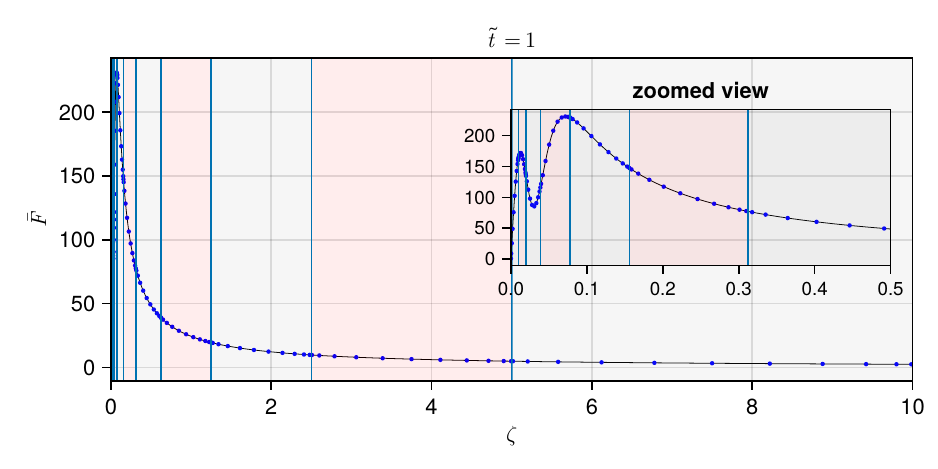}
    \caption{Representation of the discretization used in the example, with the graph of $\Bar{F}$.}
    \label{fig: F_interpolant_subintervals}
\end{figure*}
\begin{figure}[!htb]
    \centering
    \includegraphics[width=0.6\linewidth]{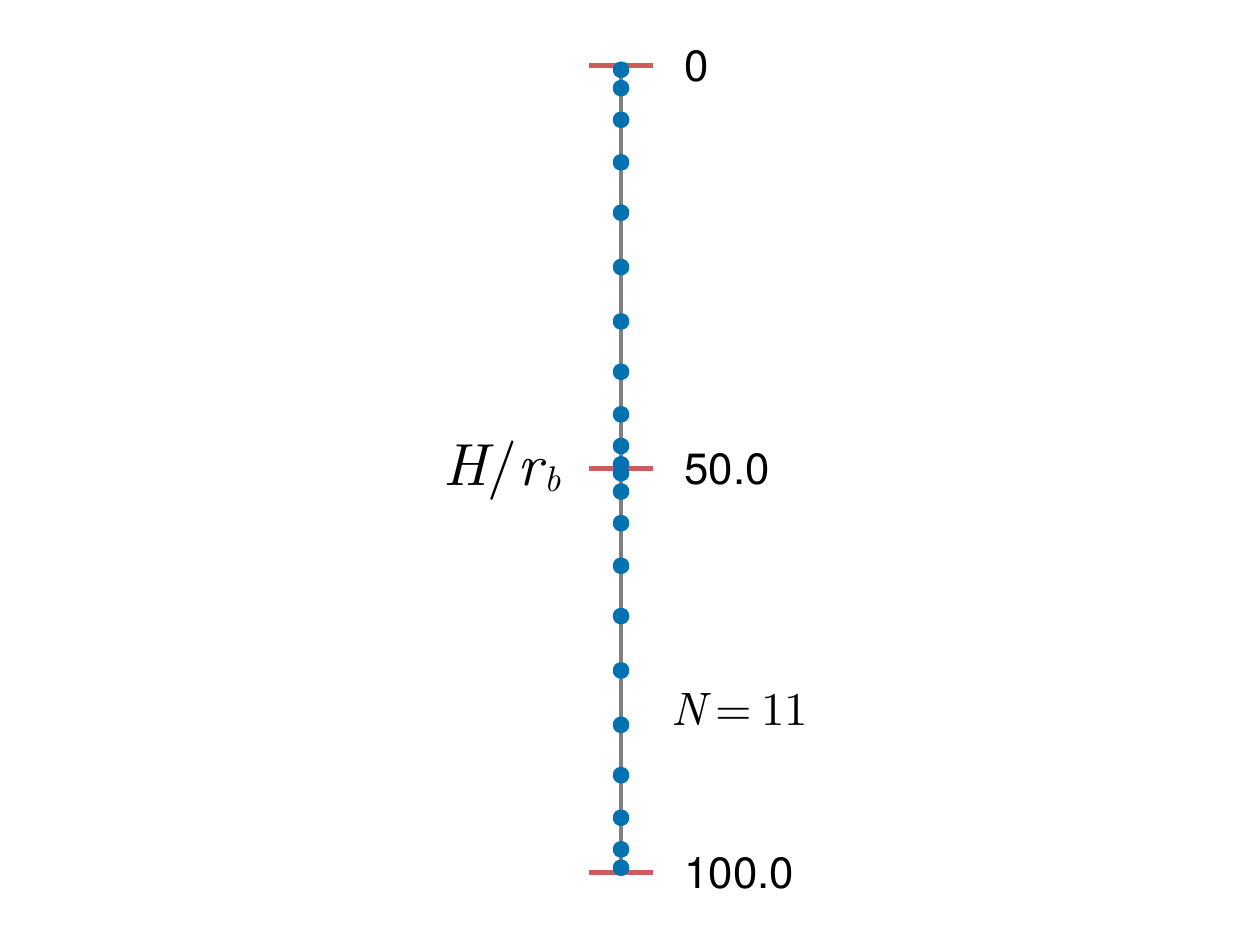}
    \caption{Representation of the subsegments used in the example. The integration nodes shown correspond to the case with least points utilized.}
    \label{fig: line_discretization}
\end{figure}

\begin{figure*}
    \textbf{Density of $\log_{10}{\epsilon}$ of the response at a point due to a point source}
    \centering
    \includegraphics[width=\linewidth]{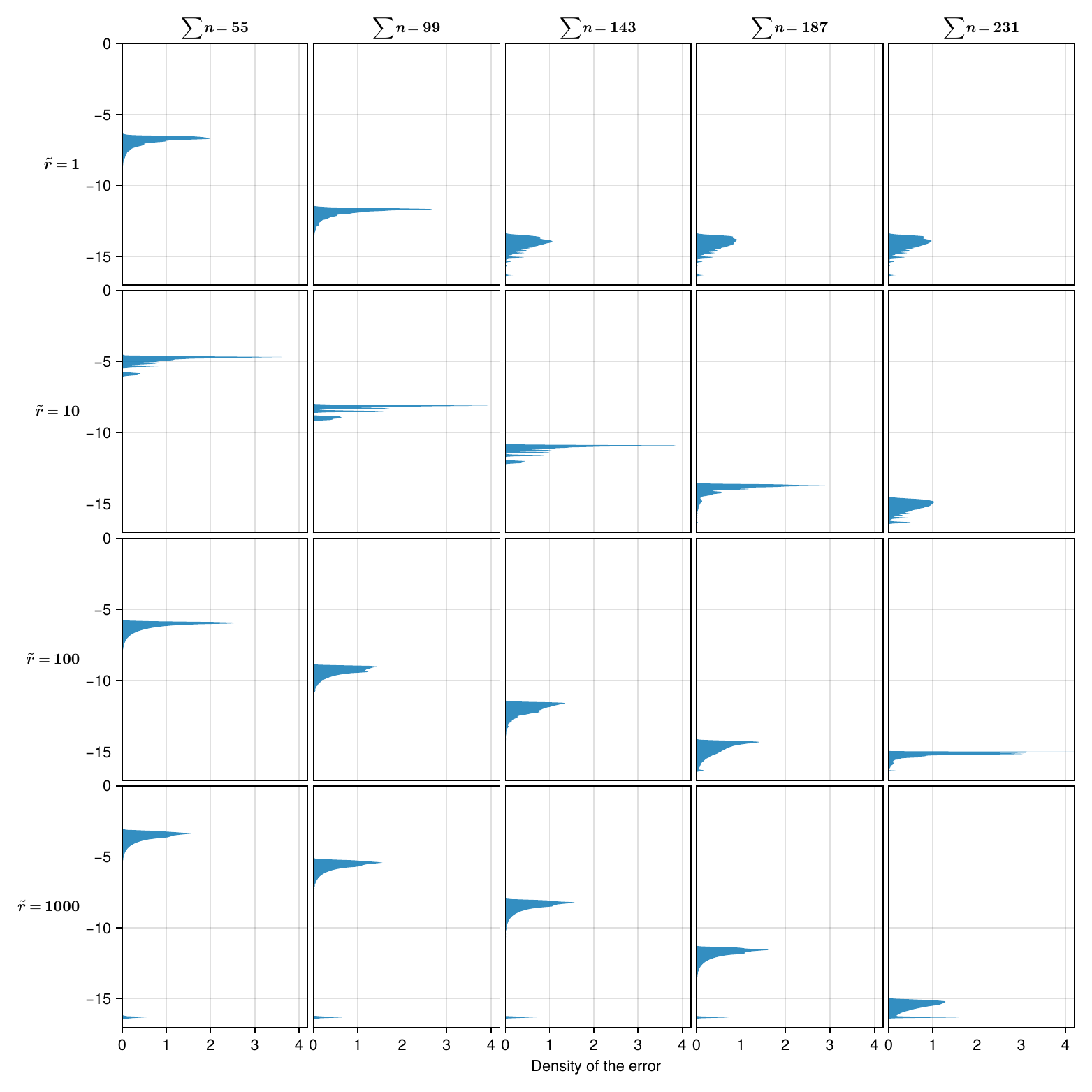}
    \caption{Density of the logarithm of the absolute error $\log_{10}(\epsilon)$ of the point source response with the load of Equation \eqref{eq: synthetic_load} during a 20 year long hourly simulation, comparing against the convolution in Equation \eqref{eq: convolution}, evaluated at a point at several values of the non-dimensional distance $\tilde{r}$ and the total number of interpolation points $\sum n$. The $11$ integration segments used are the result of the adaptive algorithm described in Appendix \ref{section: appendix_subintervals}, using a relative tolerance of $\sqrt{\varepsilon}$, where $\varepsilon$ is double precision. The amount of points $n+1$ used in each interval are $5,9,13,17,21$, respectively.}
    \label{fig: point_to_point}
\end{figure*}

We use a cutoff of $b=10$, which in our case of a hourly time step, makes the error in Equation \eqref{eq: truncated_integral_F_exp} negligible. 
As discussed in Section \ref{section: integration}, we also divide the integration region into subintervals to improve the accuracy without compromising performance. However, there are many ways to do so. We explain the details of our implementation in Appendix \ref{section: appendix_subintervals} based on an adaptive algorithm. The subintervals used in the example are shown in Figure \ref{fig: F_interpolant_subintervals}. 
The results for the error of the point to point computation are presented in Figure \ref{fig: point_to_point} for several values of $\tilde{r}$.

\begin{figure*}
    \textbf{Density of $\log_{10}{\epsilon}$ of the response at a point due to a finite line source}
     \centering
     \includegraphics[width=\linewidth]{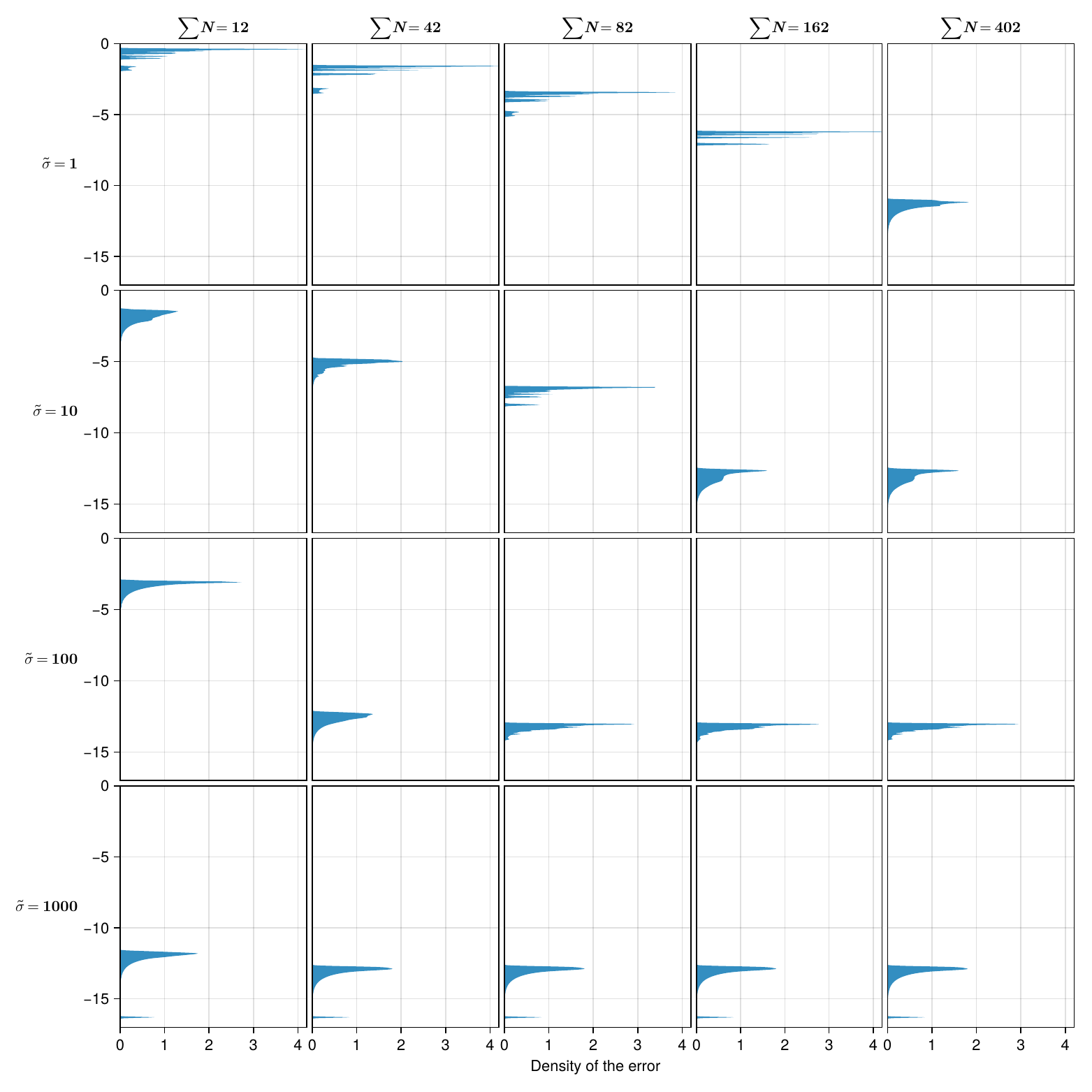}
     \caption{Density of the logarithm of the absolute error $\log_{10}(\epsilon)$ of the finite line source response with the load in Equation \eqref{eq: synthetic_load} during a 20 year long hourly simulation, comparing against the convolution in Equation \eqref{eq: convolution}, evaluated at a point at several values of the non-dimensional distance $\tilde{\sigma}$ and the total number of interpolation points $\sum N$ for the line. The evaluation point lies at $z$ equal the midpoint of the source line, of non-dimensional length $H/r_b = 1000$. A total of $\sum n = 231$ points are used to perform the integral in $\zeta$ as per the findings shown in Figure \ref{fig: point_to_point}.}
     \label{fig: segment_to_point}
\end{figure*}

We also perform a similar simulation for the response at a point due to a line source case. The line has length $\frac{H}{r_b} = 1000$ and the evaluation point lies at its mid-length, at several radial distances $\tilde{\sigma}$. In this example, we analyze how the discretization of the finite line affects the accuracy of the result. The results are shown in Figure \ref{fig: segment_to_point}. We divided the line in $2$ subsegments of equal length, as shown in Figure \ref{fig: line_discretization}. For each subsegment we applied the $(N+1)$-points Gauss-Legendre rule, as in Equation \eqref{eq: z_integral_discrete}, starting at $N=5$ up to $N=200$.

\begin{figure*}
    \textbf{Density of $\log_{10}{\epsilon}$ of the response at a finite line due to another finite line source}
     \centering
     \includegraphics[width=\linewidth]{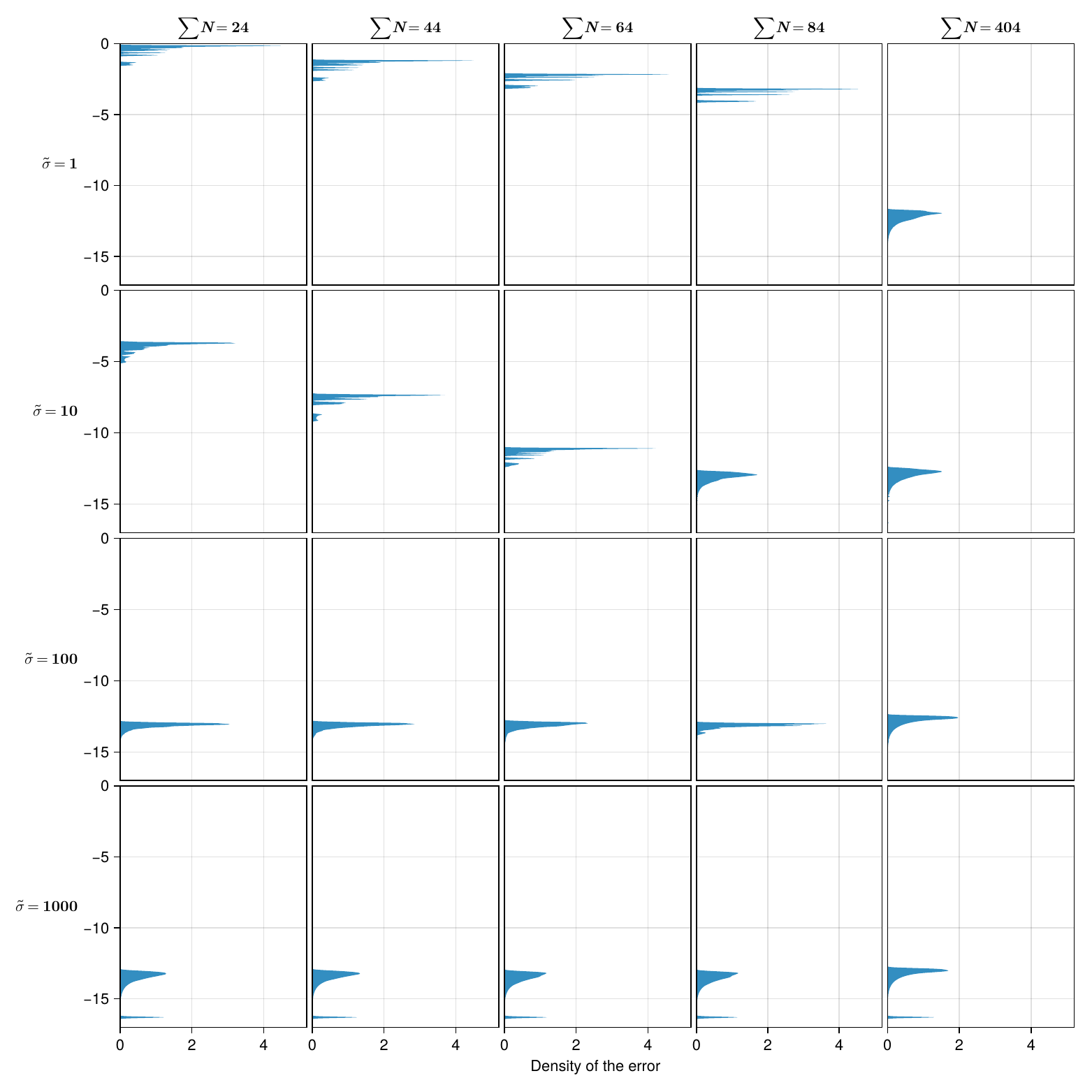}
     \caption{Density of the logarithm of the absolute error $\log_{10}(\epsilon)$ of the finite line source response with the load in Equation \eqref{eq: synthetic_load} during a 20 year long hourly simulation, comparing against the convolution in Equation \eqref{eq: convolution}, evaluating the mean response at another segment at several values of the non-dimensional distance $\tilde{\sigma}$ and the total number of interpolation points $\sum N$ for each line. Both the source and evaluation segments are at the same depth and length $H/r_b = 1000$. A total of $\sum n = 231$ points are used to perform the integral in $\zeta$ as per the findings shown in Figure \ref{fig: point_to_point}.}
     \label{fig: segment_to_segment}
\end{figure*}

Finally, a similar analysis is performed for a simulation of the mean response at a segment due to a source segment. Both segments are $\frac{H}{r_b} = 1000$, and the simulation is performed at several radial distances $\tilde{\sigma}$. The results are shown in Figure \ref{fig: segment_to_segment}. In this case, $4$ subsegments for the lines have been used, with the same amount of points $N$ and $N^\prime$ for both lines ranging from $N = 20$ to $N=400$.

In sight of the results, there are a few remarks that are interesting to make.
\begin{itemize}
    \item The error in all three cases is close to double precision for a large enough value of $\sum n$. Nevertheless, we see an increase in the number of integration points necessary to achieve good accuracy in the line to point, and especially, in the line to line cases. This is due to the extra integrals that must be performed over each lines, which contribute to accumulating error. The effect is accentuated for small values of $\tilde{\sigma}$, of the order of $1$ or less. This happens because the variation in the distance between the evaluation points and each point in the line varies more the lower the radial distance, requiring more integration points.
    \item For the $\zeta$ discretization, increasing the number of points $\sum n$ makes the error decrease. However, after a certain threshold of $\sum n$, the error does not improve anymore with more points. The same effect happens for the line discretization with $\sum N$.
    \item In the line integrals, lower values of $\tilde{\sigma}$ yield greater errors. The range of distances between points in the source and target (and hence, the integrand, which depends on $\tilde{r}$ through the vectors $R_\text{STP}$ and $ R_\text{STS}$) experiences a larger variation for smaller values of $\tilde{\sigma}$. This makes the use of subsegments very beneficial, especially for small values of $\tilde{\sigma}$.
    \item The error stays mostly stable during the duration of the simulation. As a special case, for $\tilde{r} = 1000$, at the start of the simulation the error is very low, only to grow very rapidly. This is not a mistake: it happens because before that point, the heat wave has not yet arrived at the evaluation point, and in such a situation, instead of trying to compute it we just assign it a value of $0$, since the real value is orders of magnitude below double precision.
    \item Increasing the number of points in the discretizations clearly improves the precision of the method, but comes at the expense of an increase in the computational cost. This is relevant because it is possible to run the computations faster if lower accuracy is required for a particular application. 
\end{itemize}

 
Let us now analyze the computational complexity of the method presented in Section \ref{section: integration}.
At each time step, the updating (Equation \eqref{eq: F_recursive}) and integration (Equation \eqref{eq: temperature_from_F}) operations must be performed in order to compute the temperature. 
First, note that the complexity remains constant at every time step, since according to Equation \eqref{eq: F_recursive}, the updating rule only involves the last value of $\tilde{F}$ and the last value of the load. Therefore, the complexity of the algorithm is linear in the number of time steps $N_t$.

As for the integration, we would like to remark that in all three cases of Equations \eqref{eq: integral_exp_point}, \eqref{eq: integral_exp_seg_to_point}, and \eqref{eq: integral_exp_seg_to_seg}, only $f$ changes through time. This is so because $P$, $R$, and $v$ depend on the discretizations that we choose and the evaluation distance $\tilde{r}$, while $f$ depends on the integrand $\tilde{F}_\text{exp}$, which is updated at each time step. 
This has two implications. 
First, we can \textit{precompute} $P$, $R$, $v$, and their product, obtaining as a result a vector $\lambda$ (whose expression is different for each configuration).
Then, at each time step, the integral is obtained by the dot product $\lambda \cdot f$, reducing significantly the computations required for each time step.
Additionally, this means that the actual implementation of the updating and integration steps is independent of the particular configuration we are studying.
As a consequence, the computational cost of a time step is $\mathcal{O}(n m)$ and it is the same for all three cases, where $m$ is the number of subintervals and $n$ the amount of points in each subinterval. Note that this is independent of $N$ and $N^\prime$ because the computations related to the finite line discretization are performed in the precomputation phase and have no computational impact in the update of each time step. 

Overall, we conclude that the presented algorithm has a computational complexity of $\mathcal{O}(N_t n m)$. This represents a large speedup from the $\mathcal{O}(N_t^2)$ complexity obtained by computing the temperature naïvely through the convolution.

We believe the execution time is not a good measure of performance of algorithms, since it can vary across orders of magnitude depending on its actual implementation. However, as a rough benchmark for the presented algorithm, simulations with $10^6$ time steps take a fraction of a second to run.

\section{Conclusions}
We have presented a “non-history dependent” algorithm to compute the temperature field created by a point source and a finite line source based on the works of Lamarche \cite{lamarche2007, lamarche2009}, and \cite{ours}. 
%We also explained how the method, with some minor modifications, can be also used to compute temperature fields created by a finite line source, either evaluated at a point or as the mean over another segment.

The scheme has the favorable iterative properties discussed by Lamarche, with each update only requiring information on the last temporal discretization step rather than its whole history. A key insight in the derivation was to use the inverse Laplace transform analytically of the impulse response of a point source, which yields a simple analytical expression. As a consequence, our scheme can be written in a closed, relatively simple form. 
The computation of the temperature additionally requires the evaluation of an oscillatory integral, whose oscillation frequency depends on the evaluation distance. This makes it necessary to use suitable numerical integration techniques. To solve the problem at hand, we have proposed an \textit{ad hoc} scheme based on the techniques put forward by Bakhvalov and Vasil'eva \cite{bakhvalov}, which makes use of Legendre polynomials to produce an interpolation of the integrand. This allows to approximate the integral by an expression where the physical variables appear separated. This fact is exploited to extend the method to the line source cases.

We have also checked the accuracy of the proposed method by comparing the results obtained against the numerical convolution for a synthetic hourly load with yearly and daily oscillations.
The results obtained show that by increasing the number of integration points it is possible to approach double precision accuracy in all three studied cases. However, the number of integration points needed to reach good accuracy increases in the line to point and line to line cases, especially so for small values of $\tilde{\sigma}$. This shortcoming is something we would like to mitigate in future works.
The computational complexity of the algorithm with the proposed implementation is linear in both the number of time steps and discretization points.
It is therefore possible to tune the number of integration points to balance precision and performance.
A relevant remark is that this is an improvement over the quadratic computational complexity of the convolution, allowing for finer time discretizations.

We believe that the observed performance results are very promising, and have the potential to improve the scaling properties of current techniques, enabling complex borehole network simulations.

\section*{Acknowledgments}
The authors acknowledge Energimyndigheten (Swedish Energy Agency) for financing this research via the grants P2022-00486, P2022-00499 (TERMO program) and P2022-01040 (JPP SES and GEOTHERMICA ERA-Net network programs).

\begin{appendices}

\section{Estimate of the truncation error in Equation \texorpdfstring{\eqref{eq: truncated_integral_F_exp}}{(23)}}
\label{section: appendix_error}

The aim of this section is to estimate the error committed by using approximation in Equation \eqref{eq: truncated_integral_F_exp}.
It is enough to estimate the error of computing $\int_0^b \Bar{F}(\zeta, \tilde{t}) e^{i \tilde{r} \zeta} \ \mathrm{d}\zeta$ instead of $\int_0^\infty \Bar{F}(\zeta, \tilde{t}) e^{i \tilde{r} \zeta} \ \mathrm{d}\zeta$. Then,
\begin{align*}
    \varepsilon &= \bigg|  \int_0^\infty \Bar{F}(\zeta, \tilde{t}) e^{i \tilde{r} \zeta} \ \mathrm{d}\zeta - \int_0^b \Bar{F}(\zeta, \tilde{t}) e^{i \tilde{r} \zeta} \ \mathrm{d}\zeta   \bigg| \\ 
    &=  \bigg|   \int_b^\infty \Bar{F}(\zeta, \tilde{t}) e^{i \tilde{r} \zeta} \ \mathrm{d}\zeta   \bigg| \leq  \int_b^\infty \big| \Bar{F}(\zeta, \tilde{t}) \big|   \mathrm{d}\zeta  \\
    &\leq \sum_{i=0}^{n-1} \int_b^\infty  \frac{\left(1 - e^{-\zeta^2 \Delta \tilde{t}} \right)}{\tilde{r}\zeta} e^{-\zeta^2 (n-i) \Delta \tilde{t}} \big| \tilde{q}^\prime(i \Delta \tilde{t}) \big|  \mathrm{d}\zeta \\
    &\hspace{15pt}+ \int_b^\infty \big| \tilde{q}^\prime(n \Delta \tilde{t}) \big|  \frac{ e^{-\zeta^2 \Delta \tilde{t}}}{\tilde{r}\zeta} \mathrm{d}\zeta \\
    &\leq \sum_{i=0}^{n-1}\frac{\big| \tilde{q}^\prime(i \Delta \tilde{t}) \big|}{\tilde{r} b} \int_b^\infty e^{-\zeta^2 (n-i) \Delta \tilde{t}} \mathrm{d}\zeta + \frac{\big| \tilde{q}^\prime(n\Delta \tilde{t}) \big|}{\tilde{r} b} \int_b^\infty e^{-\zeta^2  \Delta \tilde{t}} \mathrm{d}\zeta \\
    &\leq \frac{Q}{\tilde{r} b} \int_b^\infty  \sum_{i=0}^{n-1} e^{-\zeta^2 (n-i) \Delta \tilde{t}} + e^{-\zeta^2  \Delta \tilde{t}} \mathrm{d}\zeta  \\
    & = \frac{Q}{\tilde{r} b} \sqrt{\frac{\pi}{4  \Delta \tilde{t}}} \left( \sum_{i=0}^{n-1} \frac{\text{erfc} \left( b \sqrt{(n-i) \Delta \tilde{t}}\right)}{\sqrt{(n-i)}} + \text{erfc} \left( b \sqrt{\Delta \tilde{t}}\right)\right) \ ,
\end{align*}
where $Q = \max\limits_{0\leq i \leq n} \big| \tilde{q}^\prime(i \Delta \tilde{t}) \big|$. 
For large enough values of $b$, the second leading term is negligible compared to the leading term, so a good estimate is 
\begin{align*}
    \varepsilon \lesssim \frac{Q}{\tilde{r} b} \sqrt{\frac{\pi}{\Delta \tilde{t}}} \ \text{erfc} \left( b \sqrt{\Delta \tilde{t}}\right) \ .
\end{align*}
As an example, for a hourly simulation, and the values $\alpha = 10^{-6}, \ r_b = 0.1, \ \tilde{r} = 10, \ Q = 100$, a value of $b=10$ gives a bound for the error of around $6 \cdot 10^{-17}$, which is below double precision.

\section{Selection of the integration subintervals}
\label{section: appendix_subintervals}

A sensible strategy for choosing subintervals that yields good accuracy is the use of an adaptive integration method.

Recall that the accuracy of the Bakhalov and Vasil'eva \cite{bakhvalov} method for the integration of $F_\text{exp}$ depends exclusively on how accurate is the representation of $\bar{F}$ via a truncated series of Legendre polynomials, where $\bar{F}$ is the non-oscillating component of $F_\text{exp}$.
The idea of adaptive integration is to find a set of segments such that the application of the Gauss-Legendre quadrature rule for the integration of the function $\bar{F}$ to each of them yields a total error below a given threshold. The discretization obtained will ensure an appropriate representation for the function $\bar{F}$ via Legendre polynomials, and as a consequence, the accurate evaluation of the integral of $F_\text{exp}$.
Since the discretization depends only on $\bar{F}$ (Equation \eqref{eq: F_expanded}), $\tilde{r}$ only plays the role of a scaling factor, so the same integration nodes and weights can be used for any geometrical configuration.

% Doing the discretization optimization focusing on $\bar{F}$ 
%  has two advantages: we can use well-known adaptive algorithms such as the one proposed by Gauss-Kronrod, and we eliminate the dependency on $\tilde{r}$ in the estimation of the discretization.

In order to estimate the error for a given discretization, we apply the method used in Gauss-Kronrod adaptive integration \cite{kronrod}, where for each interval, the error is taken to be the difference between both Gauss-Legendre and Kronrod quadrature rules, which we chose of $8$th order. The algorithm stops when the desired error tolerance (either absolute or relative) is satisfied. We have used a relative tolerance of $\sqrt{\epsilon}$, where $\epsilon$ is the double precision. This defines the set of segments that we are going to use.
Then, for a given $n$, we take the $(n+1)$-point Gauss-Legendre quadrature rule corresponding to each of the segments to build a composite quadrature rule over the original segment. For our error analysis in the paper, we have chosen the values $n=4,8,12,16,20$, with $n=20$ yielding the best result.

Then, recalling that the $(n+1)$-point quadrature rule integrates polynomials of degree $2n+1$ exactly, one can see that the same set of segments (and thus also nodes) can also be used to approximate the integrals of $\Bar{F} (\zeta) P_k(\zeta)$, for $k=0 \ldots n$, if $n$ is large enough.
Then, by using the same set of nodes and weights, we approximate the integrals in Equation \eqref{eq: ck}, and consequently in Equation \eqref{eq: integral_approx}, to obtain a good accuracy for the oscillatory integral in Equation \eqref{eq: truncated_integral_F_exp}.

In principle, in order to attain the desired accuracy throughout the simulation, one could run the described adaptive method at each time step.
However, it is highly desirable that the points where $\Bar{F}$ is evaluated remain constant at all time steps for two reasons.
First, the computational time would increase since we would not be able not precompute the matrix $P$ in Equation \eqref{eq: integral_matricial}.
Second, since the shape of the function changes, the nodes obtained by the adaptive method could also potentially change, and we would need to compute the value of the function $\Bar{F}$ at the new nodes. For fixed nodes, this presents no problem since we have the recursive relation in Equation \eqref{eq: F_recursive}. However, for dynamic nodes, we would need to turn to the Legendre polynomials interpolation of $\Bar{F}$, which would introduce error.

In view of the problems of having a dynamical discretization, we opted for a static discretization aiming to represent the shape of $\Bar{F}$ as good as possible throughout the simulation time span.
Our choice is 
\begin{align*}
    \bar{F}_\text{guess}(\zeta) = \frac{1 - e^{-\zeta^2 \Delta \tilde{t}}}{\zeta} \sum_{i=0}^K \left( w_i e^{-T_i \zeta^2 \Delta \tilde{t}} \right) \ ,
\end{align*}
which corresponds to the superposition of $K$ load pulses over a time step, where each happened $T_i$ time steps ago with amplitude $w_i$.
We have chosen the values $T = (1, 24 \cdot 7, 24 \cdot 365, 24 \cdot 365 \cdot 20)$ and $w = (1, 10, 100, 1000)$, which in the case of hourly time steps, represent pulses that happened an hour, a week, a year, and 20 years ago, respectively. The intensities are chosen in such a way that pulses more distant in the past are still relevant in the present. 

Note that with this scheme, the selected segments only depend on the cutoff $b$ and the error tolerance, while the nodes and weights also depend on the value $n$. 
The effect of using the described algorithm is that the subsegments are more densely placed in the regions where the integrand function changes rapidly.
\end{appendices}

\section*{Nomenclature}

\begin{tabular}{c c l}
    $\alpha$ & = & Thermal diffusivity (m$^2$/s) \\
    $k_g $ & = & Thermal conductivity (W/(m K)) \\
    $q^\prime$ &= & Heat transfer rate. (W/m) for linear sources \\
     & & and (W) for point sources \\
    $\tilde{q}^\prime$  & = & Reparametrized $q^\prime$ with the variable $\tilde{t}$. \\
     & & (W/m) for linear sources and (W) for point sources \\
    $T$ & = & Temperature of the medium ($^\circ$C) \\
    $T_0$ & = & Initial temperature of the medium ($^\circ$C) \\
    $r$ & = & Radial distance from the point source (m) \\
    $r_b$ & = & Scaling distance equivalent to the borehole radius (m) \\
    $\tilde{r}$ & = & non-dimensional distance $r/r_b$ from the line source (-)\\
    $t$ & = & time ($s$) \\
    $\tilde{t}$ & = & non-dimensional time $\alpha t /r_b^2$ (-) \\
    $J_k$ & = & Bessel function of the first kind of order $k$ \\
    $P_k$ & = & Legendre polynomial of order $k$ \\
    $\frac{dh}{dt}$ & = & Impulse response\\
    $F$ & = & Temporal factor of the non-history method
\end{tabular}

\subsection*{Subscripts}
\begin{tabular}{c c l}
$g$ & = & ground \\
$b$ & = & borehole \\
$\text{PS}$ & = & point source \\
$\text{STP}$ & = & segment source to target point \\
$\text{STS}$ & = & segment source to target segment \\
\end{tabular}

\printbibliography

@article{MARCOTTE2008651,
title = {Fast fluid and ground temperature computation for geothermal ground-loop heat exchanger systems},
journal = {Geothermics},
volume = {37},
number = {6},
pages = {651-665},
year = {2008},
issn = {0375-6505},
doi = {https://doi.org/10.1016/j.geothermics.2008.08.003},
url = {https://www.sciencedirect.com/science/article/pii/S0375650508000552},
author = {Denis Marcotte and Philippe Pasquier},
}

@article{ClaJav2012,
  title={A load-aggregation method to calculate extraction temperatures of borehole heat exchangers},
  author={Claesson, Johan and Javed, Saqib},
  journal={ASHRAE Transactions},
  volume={118},
  number={1},
  pages={530--539},
  year={2012}
}

@article{MitSpi2019,
author = {Matt S. Mitchell and Jeffrey D. Spitler},
title = {Characterization, testing, and optimization of load aggregation methods for ground heat exchanger response-factor models},
journal = {Science and Technology for the Built Environment},
volume = {25},
number = {8},
pages = {1036-1051},
year  = {2019},
publisher = {Taylor & Francis},
doi = {10.1080/23744731.2019.1648936},
}

@article{Bernier2004,
author = {Michel Bernier and Patrice Pinel and Richard Labib and Raphaël Paillot},
title = {A Multiple Load Aggregation Algorithm for Annual Hourly Simulations of GCHP Systems},
journal = {HVAC\&R Research},
volume = {10},
number = {4},
pages = {471-487},
year = {2004},
publisher = {Taylor & Francis},
doi = {10.1080/10789669.2004.10391115},
URL = {https://doi.org/10.1080/10789669.2004.10391115},
}

@article{LAZZAROTTO2014265,
title = {A network-based methodology for the simulation of borehole heat storage systems},
journal = {Renewable Energy},
volume = {62},
pages = {265-275},
year = {2014},
issn = {0960-1481},
doi = {https://doi.org/10.1016/j.renene.2013.07.020},
url = {https://www.sciencedirect.com/science/article/pii/S0960148113003650},
author = {Alberto Lazzarotto},
}

@article{lamarche2007,
title = {A fast algorithm for the simulation of GCHP systems},
author = {Lamarche, Louis and Beauchamp, Benoit},
year = {2007},
month = {01},
pages = {470-476},
volume = {113},
journal = {ASHRAE Transactions}
}

@article{lamarche2009,
    title = {A fast algorithm for the hourly simulations of ground-source heat pumps using arbitrary response factors},
    journal = {Renewable Energy},
    volume = {34},
    number = {10},
    pages = {2252--2258},
    year = {2009},
    issn = {0960-1481},
    doi = {https://doi.org/10.1016/j.renene.2009.02.010},
    url = {https://www.sciencedirect.com/science/article/pii/S0960148109000731},
    author = {Lamarche, Louis},
}

@book{oberhettinger2012tables,
  title={Tables of Laplace Transforms},
  author={Oberhettinger, F. and Badii, L.},
  isbn={9783642656453},
  year={2012},
  publisher={Springer Berlin Heidelberg}
}

@article{bakhvalov,
    title = {Evaluation of the integrals of oscillating functions by interpolation at nodes of gaussian quadratures},
    journal = {USSR Computational Mathematics and Mathematical Physics},
    volume = {8},
    number = {1},
    pages = {241--249},
    year = {1968},
    issn = {0041-5553},
    doi = {https://doi.org/10.1016/0041-5553(68)90016-5},
    url = {https://www.sciencedirect.com/science/article/pii/0041555368900165},
    author = {Bakhvalov, N.S. and Vasil'eva, L.G.}
}

@article{evans1999,
    title = {A comparison of some methods for the evaluation of highly oscillatory integrals},
    journal = {Journal of Computational and Applied Mathematics},
    volume = {112},
    number = {1},
    pages = {55--69},
    year = {1999},
    issn = {0377-0427},
    doi = {https://doi.org/10.1016/S0377-0427(99)00213-7},
    author = {G.A. Evans and J.R. Webster},
}

@article{CIMMINO2014641,
title = {A semi-analytical method to generate g-functions for geothermal bore fields},
journal = {International Journal of Heat and Mass Transfer},
volume = {70},
pages = {641-650},
year = {2014},
issn = {0017-9310},
doi = {https://doi.org/10.1016/j.ijheatmasstransfer.2013.11.037},
url = {https://www.sciencedirect.com/science/article/pii/S0017931013009915},
author = {Massimo Cimmino and Michel Bernier},
}

@article{LAMARCHE2007188,
title = {A new contribution to the finite line-source model for geothermal boreholes},
journal = {Energy and Buildings},
volume = {39},
number = {2},
pages = {188-198},
year = {2007},
issn = {0378-7788},
doi = {https://doi.org/10.1016/j.enbuild.2006.06.003},
url = {https://www.sciencedirect.com/science/article/pii/S0378778806001824},
author = {Louis Lamarche and Benoit Beauchamp},
}

@article{claesson2011analytical,
  title={An Analytical Method to Calculate Borehole Fluid Temperatures for Time-scales from Minutes to Decades.},
  author={Claesson, Johan and Javed, Saqib},
  journal={ASHRAE Transactions},
  volume={117},
  number={2},
  year={2011}
}

@book{carslaw1959conduction,
  title={Conduction of Heat in Solids},
  author={Carslaw, H.S. and Jaeger, J.C.},
  isbn={9780198533689},
  lccn={85026963},
  series={Oxford science publications},
  year={1959},
  publisher={Clarendon Press}
}

@phdthesis{eskilson1987,
title = "Thermal analysis of heat extraction boreholes",
author = "Per Eskilson",
year = "1987",
language = "English",
isbn = "91-7900-298-6",
type = "Doctoral Thesis",
school = "Department of Physics",
}

@article{ours,
  title={A non-history dependent temporal superposition algorithm for the point source solution},
  author={Lazzarotto, Alberto and Basquens, Marc and Cimmino, Massimo},
  journal={IGSHPA Research Conference Proceedings},
  year={2024},
doi={https://doi.org/10.22488/okstate.24.000021}
}

@article{Yavuzturk_Jeffrey,
title={ A short time step response factor model for vertical ground loop heat exchangers },
author={ C. Yavuzturk and Jeffrey D. Spitler },
year={ 1999 },
month = {7},
publisher={ American Society of Heating, Refrigerating and Air-Conditioning Engineers, Inc., Atlanta, GA (US) },
url = {https://www.osti.gov/biblio/20085638},
journal={ASHRAE Transactions},
volume={105(2)},
pages={475--485}
}

@article{kronrod,
title={Integration with control of accuracy},
author={A. S. Kronrod},
year={ 1964 },
journal={Dokl. Akad. Nauk SSSR},
volume={154},
issue={2},
pages={283--286}
}

@article{sts_steady_state,
author = {Cimmino, Massimo and Cook, Jonathan C. and Isiordia Farrera, José A. },
title = {Optimal discretization of geothermal boreholes for the calculation of g-functions},
journal = {Science and Technology for the Built Environment},
volume = {30},
number = {3},
pages = {234--249},
year = {2024},
publisher = {Taylor \& Francis},
doi = {10.1080/23744731.2023.2295823},
URL = {  https://doi.org/10.1080/23744731.2023.2295823},
eprint = {   https://doi.org/10.1080/23744731.2023.2295823 }
}

@article{LAMARCHE2017466,
title = {Mixed arrangement of multiple input-output borehole systems},
journal = {Applied Thermal Engineering},
volume = {124},
pages = {466-476},
year = {2017},
issn = {1359-4311},
doi = {https://doi.org/10.1016/j.applthermaleng.2017.06.060},
url = {https://www.sciencedirect.com/science/article/pii/S1359431117307500},
author = {Louis Lamarche},
}

\end{document}